\documentclass{elsart}

\usepackage{natbib}

\usepackage{graphicx}
\usepackage{epsfig}

\usepackage{amssymb}

\begin{document}

\begin{frontmatter}



\title{Chandra X-ray Observations of Newly Discovered, $z$ $\sim$ 1 Clusters from the Red-Sequence Cluster Survey\thanksref{titlefn}}
\thanks[titlefn]{COSPAR conference attendance was partially supported by an International Travel Grant from the American Astronomical Society and the National Science Foundation.}



\author[CU]{A.K. Hicks}\ead{amalia.hicks@colorado.edu}, 
\author[CU]{E. Ellingson}, 
\author[MIT]{M. Bautz}, 
\author[TOR]{H.K.C. Yee}, 
\author[CAR]{M. Gladders}, 
\author[PENN]{G. Garmire}
\address[CU]{Center for Astrophysics and Space Astronomy, University of Colorado at Boulder, Campus Box 389, Boulder, CO 80309, USA}
\address[MIT]{MIT Center for Space Research, 77 Massachusetts Ave., Cambridge, MA 02139, USA}
\address[TOR]{Department of Astronomy and Astrophysics, University of Toronto, 60 St. George St., Toronto, ON, M5S 3H8, Canada}
\address[CAR]{Carnegie Observatories, 813 Santa Barbara St., Pasadena, CA, 91101, USA}
\address[PENN]{Department of Astronomy and Astrophysics, 525 Davey Lab, The Pennsylvania State University, University Park, PA, 16802, USA}

\begin{abstract}
Observational studies of cluster evolution over moderate redshift
ranges (to $z$ $\sim$ 1) are a powerful tool for constraining
cosmological parameters, yet a comprehensive knowledge of the
properties of these clusters has been hitherto unattained.  Using a
highly efficient optical selection technique, the Red-Sequence Cluster
Survey (RCS) has unearthed a large sample of high redshift cluster
candidates.  All six of the clusters from this sample which have been
observed with the Chandra X-Ray Observatory were detected in the
X-ray.  These Chandra follow-up observations (0.64 $<$ $z$ $<$ 1.0)
indicate that the clusters are systematically less luminous than their
similarly rich, X-ray selected counterparts at lower redshifts, though
they are consistent with standard $L_x-T_x$ relationships.
Comparisons with X-ray selected samples suggest that the discrepancy
may be due in part to systematic differences in the spatial structure
of the X-ray emitting gas.  Our initial results from Chandra follow-up
observations of six RCS clusters are presented, including $\beta$
model parameters and spectral information.    
\end{abstract}


\end{frontmatter}



\section{The RCS Survey}

\begin{figure}
\includegraphics*[width=3in]{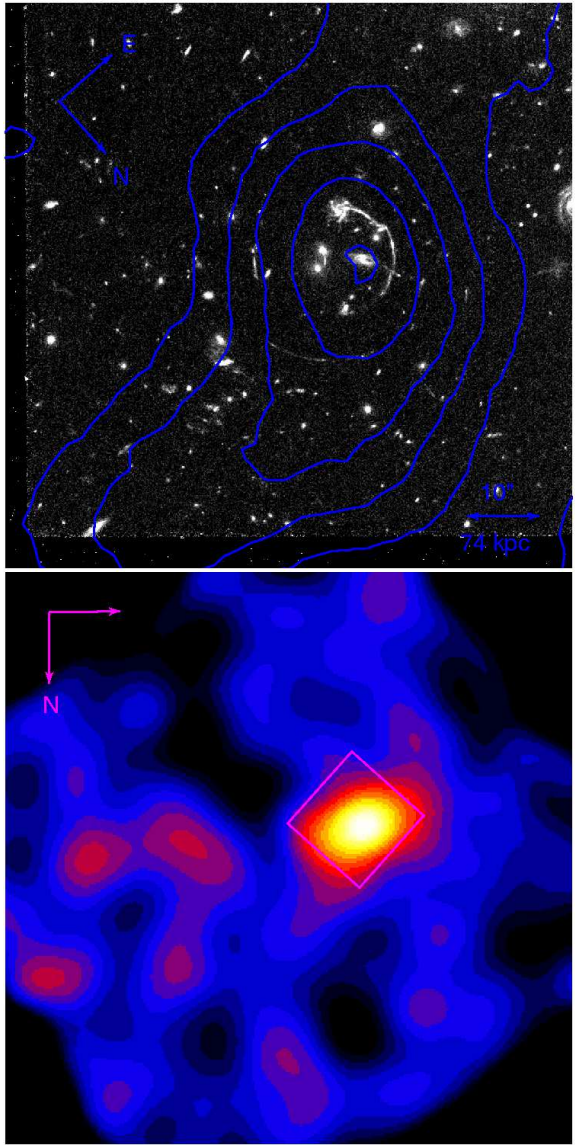}
\label{figure1}
\caption{Top: HST image of RCS 0224-0002 spectroscopically confirmed
  at $z$=0.773, with overlayed X-ray contours.  The outer
  gravitational lensing arc has been spectroscopically confirmed at a
  redshift of $z$=4.87 \citep{gladders3}, with the others expected to
  fall within the range 1.8$<$ $z$ $<$3.6  The five linearly spaced
  contours indicate values between $9.4\times10^{-6}$ and
  $2.1\times10^{-5}$ counts/$\rm{pix}^2$/$\rm{cm}^2$ and were created
  using a Gaussian smoothed (5 pixel FWHM) 0.29-7.0 keV Chandra flux image.
  Bottom: Adaptively smoothed flux image of RCS 0224-0002 created with
  the CIAO tool csmooth.  The boxed area indicates the region shown in
  the upper panel.}   
\end{figure}

The Red-Sequence Cluster Survey \citep{gladders} is a 90 square degree
optical survey performed at CFHT and CTIO using $R_C$ and $z$'
filters.  It utilizes the red sequence of elliptical galaxies to find
galactic overdensities on the sky \citep{gladders2}.  The color
information also guards against projection effects, and provides
photometric redshift estimates.  The survey has identified
between 3500 and 4000 clusters total (in the redshift range 0.2 $<$ $z$
$<$ 1.2), over 1500 of which are at least as optically rich as Abell class
0 clusters \citep{gladders}.  Simulations indicate that the survey is
complete to Abell Class 1 at a redshift of 1 for blue fractions less
than 0.45, and that false detection rates are less than 5\%, which
is significantly lower than that of single-passband optical cluster
surveys \citep{postman, donahue2,  gilbank}.  Figure \ref{figure1}
shows an example of a $z$=0.773 RCS cluster.   


\section{The Chandra Subsample}

So far we have observed six high redshift (0.64 $<$ z $<$ 1.0)
cluster candidates with the Chandra X-ray Observatory.  Table
\ref{table1} lists our current sample of follow-up observations.
All six clusters were detected in the X-ray, using the CIAO tool
csmooth, to better than $3\sigma$.  The X-ray centroids of these
clusters were all found within 10 arcseconds of their respective
optical centers.   

Spectra were extracted from 300 kpc ($\Lambda$CDM with $\rm{H}_0=70$,
$\Omega_{\rm{m}}=0.3$, and $\Omega_\Lambda=0.7$) radius regions
around the X-ray centroids of these clusters, with backgrounds taken
from each respective aimpoint chip.  The spectra were fit in XSPEC
using absorbed single temperature models with galactic column
densities and abundances fixed at 0.3 solar.  Five of the six
observations yielded enough counts to constrain a temperature.
Results are listed in Table \ref{table2}. 

\section{Optical and X-ray Properties}

\subsection{Optical Richness}



$B_{gc}$, explained in detail in \citet{bgc}, is a parameter which
describes the optical richness of a cluster of galaxies.  Technically
it is the galaxy-cluster spatial covariance amplitude, but in essence
it is simply a measure of galaxy overdensity within a given aperture,
normalized for the expected spatial distribution of galaxies in the
cluster and the evolving galaxy luminosity function.  All of the RCS
clusters that have been observed with Chandra have richnesses that
imply Abell richness classes of at least 1.  

There exist a few challenges in the calculation of $B_{gc}$ at high
redshift.  One is the lack of a complete knowledge of the galaxy
luminosity function at these redshifts.  Another is uncertainty due to 
cluster galaxy evolution.  The latter uncertainty can be minimized by
employing the parameter $B_{gc,red}$, which is essentially the same as
$B_{gc}$, but is calculated using only galaxies in the red-sequence.
Throughout this paper we will use only the parameter $B_{gc,red}$   

\begin{figure}
\includegraphics*[height=3in,angle=90]{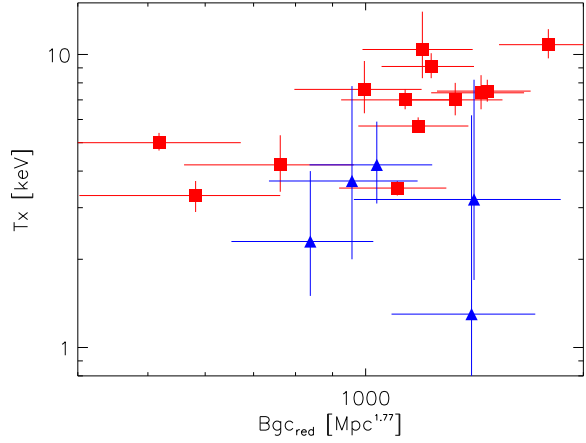}
\label{figure2}
\caption{$T_x$ vs. $B_{gc,red}$.  Each square represents one of 13
  moderate redshift (0.1 $<$ $z$ $<$ 0.6) CNOC clusters \citep{cnoc} taken
  from the Chandra archive, and triangles denote RCS clusters.
  Temperatures were calculated using spectra extracted within a 300
  kpc ($\Lambda$CDM) radius region and fit over the 0.6-7.0 keV energy
  range.  A single temperature model was used, with galactic
  absorption.  The abundance of RCS clusters was fixed at 0.3.
  Temperature error bars show 90\% confidence intervals, and
  $B_{gc,red}$ error bars are shown at 1$\sigma$.}       
\end{figure}

It is expected that $B_{gc}$ should correlate strongly with X-ray
temperature for relaxed clusters, and a trend has been seen when
optical richness is plotted versus temperature for moderate redshift
($z \sim 0.3$), X-ray selected clusters \citep{yee}.  However,
the same correlation is not apparent when RCS clusters are added to
the plot (Figure \ref{figure2}).  

In comparison with lower redshift clusters, high redshift
optically selected clusters with similar $B_{gc}$ values appear systematically
cooler and less luminous than their X-ray selected counterparts,
possibly suggesting that a smaller fraction of the intra-cluster gas
in these objects has collapsed and become virialized.  This
interpretation may seem intuitive, given that X-ray cluster surveys
preferentially select more relaxed clusters with deep potential wells
and oftentimes cooling cores.  Our findings reinforce the need to
question what defines a cluster, and whether X-ray selected clusters
primarily represent a highly virialized, high X-ray luminosity tail of
the cluster distribution.   

\subsection{X-ray Surface Brightness}

\begin{figure}
\includegraphics*[height=3in,angle=90]{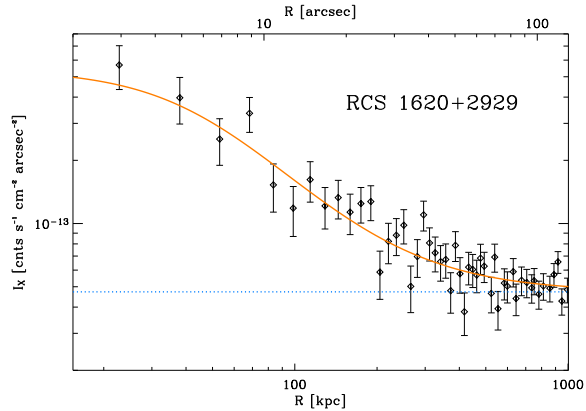}
\label{figure3}
\caption{Surface brightness was calculated for the 0.29-7.0 keV
  band in 2 arcsecond radial bins.  The solid line represents the best
  fitting $\beta$ model, and the dotted line indicates the
  fit-determined background value.}
\end{figure}

A radial surface brightness profile was computed over the range
0.29-7.0 keV in circular annuli for each cluster.  We were able to
constrain $\beta$ models for four clusters.  Best fit
parameters are expressed in Table \ref{table2} and an example is shown
in Figure \ref{figure3}.  The results of these fits are interesting in
that $\beta$ values seem systematically low for these clusters, with
the implication that, on average, high redshift optically selected
clusters are less centrally condensed than X-ray selected samples.   

\subsection{The $L_x-T_x$ Relationship}

\begin{figure}
\includegraphics*[height=3in,angle=90]{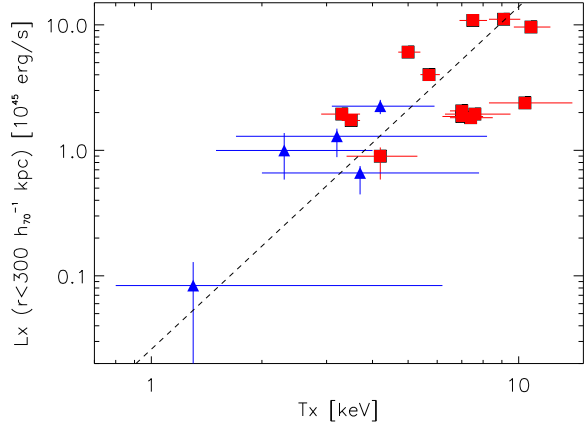}
\label{figure4}
\caption{$L_x$ vs. $T_x$.  Squares indicate 13 CNOC clusters
  and triangles represent the 5 RCS clusters.  Unabsorbed luminosities were
  calculated with XSPEC and converted to bolometric X-ray luminosities
  with PIMMS.  Temperatures were obtained with the same method as in
  Figure \ref{figure2}.  The dashed line indicates a standard power law
  with slope 2.88 \citep{arnaud}}
\end{figure}

Though many of our results indicate that RCS clusters differ
significantly from X-ray selected clusters, their X-ray properties are
consistent with a standard $L_x-T_x$ relationship.  Figure \ref{figure4}
is a plot of $L_x$ vs. $T_x$ for 13 CNOC and 5 RCS clusters.  The
consistency of RCS clusters with the $L_x-T_x$ relationship of moderate
redshift X-ray selected clusters implies that though RCS clusters may
possess relatively low mass virialized cores of gas, the gas is in a 
similar physical state to that found in X-ray selected samples.  

\section{Summary and Discussion}

The Red-Sequence Cluster Survey has uncovered a large sample of high
redshift cluster candidates.  All of the RCS clusters observed
with Chandra were detected at greater than 3$\sigma$ significance in
the X-ray.  This is evidence that the RCS method is a reliable way to
detect high redshift clusters.  The X-ray properties of RCS clusters
are consistent with a standard $L_x-T_x$ relationship.  This leads us to
believe that the X-ray gas in these clusters is in a physical state
very similar to that found in X-ray selected clusters.  

The results of detailed X-ray analysis imply that though these
clusters have extended structures of galaxies, they possess relatively
small virialized cores.  These findings are similar to those from a
number of previously conducted optical surveys \citep{donahue2, lubin,
  gilbank}.  Optical, high redshift cluster surveys regularly find
candidates with lower X-ray luminosities than those of X-ray selected
clusters.  Our study suggests that the RCS survey is detecting a
different population of clusters from that found in X-ray selected
samples, possibly including ultimately very rich clusters which are
currently in the early stages of virialization.  Similar objects have
recently been described by \citet{ford}.  Such systems may be
expected to be more common at high redshifts in a low matter density
universe.  Optical surveys are much more sensitive to
such systems than are X-ray surveys, especially if any related
filamentary structures lie along the line of sight (though it is
unlikely that all of the RCS cluster candidates that were
followed-up with Chandra are associated with such structures).    


Upcoming Chandra observations of five new and two previously observed
RCS clusters in the next AO, along with ongoing velocity dispersion
measurements and weak lensing analysis should help to provide more
definitive constraints on the masses, dynamical states, and gas
content of high redshift optically selected samples of galaxy
clusters. 



\begin{table}
\caption{Chandra Observed RCS Cluster Sample}
\label{table1}
\begin{tabular}{lccccccc}
\hline
Cluster & $z$  & Array & Exposure & Notes \\
\hline
\hline
RCS 0224-0002 &  0.77  & ACIS-S & 12620 & gravitational lens \\
RCS 0439-2904 &  0.95 & ACIS-S & 77905 & {photometric redshift} \\
RCS 1326+2903 &  0.95  & ACIS-S & 63590 & {photometric redshift} \\
RCS 1417+5305 &  1.0 & ACIS-I & 62820 & {photometric redshift} \\
RCS 1419+5326 &  0.64  & ACIS-S & 9910 & gravitational lens \\
RCS 1620+2929 &  0.87  & ACIS-S & 36640 & gravitational lens? \\
\end{tabular}  
\end{table}

\begin{table}
\caption{Optical/X-ray Properties}
\label{table2}
\begin{tabular}{lcccccccc}
\hline
Cluster & $z$ & {$B_{gc,red}$}& {${T_x}$} & {${\rm{r_c}}$} & {${\rm{r_c}}$} & {${\beta}$} \\
{} & {} & {[$\rm{Mpc}^{1.77}$]}  & {[keV]}& {[arcsec]}&{[$\rm{h}_{70}^{-1}$ kpc]} & {} \\
\hline
\hline
RCS 0224-0002 & 0.77  &${838}^{+186}_{-186}$& ${2.3}^{+1.7}_{-0.8}$ &${10}^{+2}_{-2}$& ${73}^{+15}_{-12}$& ${0.41}^{+0.07}_{-0.05}$  \\
RCS 0439-2904 & 0.95 & ${1412}^{+449}_{-449}$&  ${3.2}^{+5.0}_{-1.5}$ &${8}^{+2}_{-1}$&${66}^{+13}_{-11}$& ${0.41}^{+0.08}_{-0.05}$  \\
RCS 1326+2903 & 0.95  &${1401}^{+315}_{-315} $& ${1.3}^{+4.9}_{-0.5}$  & {...}& {...} & {...}  \\
RCS 1417+5305 &1.0   &${1325}^{+362}_{-362}$&  {...} & {...} & {...} & {...}   \\
RCS 1419+5326 & 0.64  &${1036}^{+200}_{-200}$& ${4.2}^{+1.7}_{-1.1}$   &${11}^{+2}_{-2}$&${74}^{+15}_{-12}$& ${0.67}^{+0.1}_{-0.07}$   \\
RCS 1620+2929 & 0.87  &${957}^{+222}_{-222}$& ${3.7}^{+4.1}_{-1.7}$   &${6}^{+1}_{-1}$&${45}^{+9}_{-8}$& ${0.44}^{+0.07}_{-0.05}$  \\
\end{tabular}
\end{table}


\begin{thebibliography}{}


\bibitem[Arnaud and Evrard(1999)]{arnaud} Arnaud, M, Evrard, A.E. The $L_x-T$ relation and intracluster gas fractions of X-ray clusters. MNRAS 305, 631-640, 1999.

\bibitem[Donahue et al.(2002)]{donahue2} Donahue, M., Scharf, C., Mack, J. et al. Distant cluster hunting. II. A comparison of X-ray and optical cluster detection techniques and catalogues from the ROSAT optical X-ray survey. ApJ 569, 689-719, 2002.

\bibitem[Ford et al.(2004)]{ford} Ford, H., Postman, M., Blakeslee,
  J.P. et al. The Evolutionary Status of Clusters of Galaxies at
  z~1. astro-ph/0408165, 2004.

\bibitem[Gilbank et al.(2004)]{gilbank} Gilbank, D.G., Bower, R.G.,
  Castander, F.J. et al. Exploring the selection of galaxy clusters
  and groups: an optical survey for X-ray dark clusters. MNRAS 348,
  551-580, 2004. 


\bibitem[Gladders and Yee(2004)]{gladders} Gladders, M.D., Yee,
  H.K.C. The Red-Sequence Cluster Survey I: The Survey and Cluster
  Catalogs for Patches RCS0926+37 and RCS1327+29. ApJS, 2004, in
  press. 

\bibitem[Gladders and Yee(2000)]{gladders2} Gladders, M.D., Yee, H.K.C. A new method for galaxy cluster detection. I. The algorithm. AJ 120, 2148-2162, 2000.

\bibitem[Gladders, Yee, and Ellingson(2002)]{gladders3} Gladders,
  M.D., Yee, H.K.C., Ellingson, E. Discovery of a $z$ = 0.77 galaxy
  cluster with multiple, bright, strong-lensing arcs. AJ 123, 1-9, 2002.

\bibitem[Lubin, Mulchaey and Postman(2004)]{lubin} Lubin, L.M., Mulchaey, J.S., Postman, M. The first detailed X-ray observations of high-redshift, optically selected clusters: XMM-Newton results for Cl 1324+3011 at $z$=0.76 and Cl 1604+4304 at $z$=0.90. ApJL 601, 9-12, 2004.

\bibitem[Postman et al.(1996)]{postman} Postman, M, Lubin, L.M., Gunn, J.E. et al. The Palomar Distant Clusters Survey. I. The Cluster Catalog. AJ, 111, 615-641, 1996.

\bibitem[Yee and Ellingson(2003)]{yee} Yee, H.K.C., Ellingson, E. Correlations of richness and global properties in galaxy clusters. ApJ 585, 215-226, 2003.

\bibitem[Yee, Ellingson and Carlberg(1996)]{cnoc} Yee, H.K.C., Ellingson, E., Carlberg, R.G. The CNOC cluster redshift survey catalogs. I. Observational strategy and data reduction techniques. ApJS 102, 269-287, 1996.

\bibitem[Yee and Lopez-Cruz(1999)]{bgc} Yee, H.K.C., Lopez-Cruz, O. A
  quantitative measure of the richness of galaxy clusters. AJ 117,
  1985-1994, 1999. 

\end{thebibliography}
\end{document}